\pdfoutput=1

\documentclass[10pt,aps,prl,twocolumn,superscriptaddress,footinbib]{revtex4-1}

\pdfoutput=1
\usepackage{graphicx}
\usepackage{amssymb}
\usepackage{moresize}
\usepackage{amsmath}
\usepackage{bm}
\usepackage{placeins}

\usepackage{times}
\usepackage{newtxmath}

\hbadness=\maxdimen
\vbadness=\maxdimen
\usepackage{titlesec}
\usepackage[usenames]{color}

\def\be{\begin{equation}}
\def\ee{\end{equation}}
\def\bea{\begin{eqnarray}}
\def\eea{\end{eqnarray}}
\def\ba{\begin{array}}
\def\ea{\end{array}}

\begin{document}

\title{Nonlinear Poisson effect governed by mechanical critical transition}


\author{Jordan L. Shivers}
\affiliation{Department of Chemical and Biomolecular Engineering, Rice University, Houston, TX 77005}
\affiliation{Center for Theoretical Biological Physics, Rice University, Houston, TX 77030}
\author{Sadjad Arzash}
\affiliation{Department of Chemical and Biomolecular Engineering, Rice University, Houston, TX 77005}
\affiliation{Center for Theoretical Biological Physics, Rice University, Houston, TX 77030}
\author{F.\ C. MacKintosh}
\affiliation{Department of Chemical and Biomolecular Engineering, Rice University, Houston, TX 77005}
\affiliation{Center for Theoretical Biological Physics, Rice University, Houston, TX 77030}
\affiliation{Departments of Chemistry and Physics and Astronomy, Rice University, Houston, TX 77005}



\begin{abstract}
Under extensional strain, fiber networks can exhibit an anomalously large and nonlinear Poisson effect accompanied by a dramatic transverse contraction and volume reduction for applied strains as small as a few percent. We demonstrate that this phenomenon is controlled by a collective mechanical phase transition that occurs at a critical uniaxial strain that depends on network connectivity. This transition is punctuated by an anomalous peak in the apparent Poisson's ratio and other critical signatures such as diverging nonaffine strain fluctuations.

\end{abstract}
\maketitle

When an elastic body is subjected to an infinitesimal strain $\varepsilon_\parallel$ along one axis, the corresponding strain $\varepsilon_\perp$ in the transverse direction(s) defines Poisson's ratio $\nu = -\varepsilon_\perp/\varepsilon_\parallel$ \cite{Poisson1827, landau}. Although this ratio is constrained to the range $\nu\in[-1,1/2]$ for isotropic materials in 3D, there have been numerous recent reports of anomalously large apparent Poisson's ratios exceeding $1/2$ in a variety of fibrous materials at small strain, including felt \cite{Kabla2007b} and networks of collagen \cite{Vader2009, Roeder2009, Lake2011,Steinwachs2015, Ban2019} and fibrin \cite{Brown2009, Steinwachs2015}. This corresponds to an anomalous reduction in volume under extension, in apparent stark contrast to the linear behavior of all isotropic materials, which strictly maintain or increase their volume under infinitesimal extension. This is even true of auxetic materials with $\nu<0$ \cite{Greaves2011, Rens2019, Reid2018}.   
A volume reduction under uniaxial extension can have dramatic effects in living tissue, such as the development of highly aligned, stiffened network regions with reduced porosity between contractile cells in the extracellular matrix \cite{Vader2009, Wang2015, Abhilash2014, Ban2019}.  Although it has been argued that this effect is related to stiffening and other nonlinear phenomena in such networks \cite{Ban2019,Picu2018, Kabla2007b}, it remains unclear to what extent this anomaly is controlled by network architecture and filament properties.

Here, we show that the anomalous Poisson's ratio of fiber networks is governed by a mechanical phase transition induced by applied axial strain. Using simulations of disordered networks in 2D and 3D, we show that this phenomenon is critical in nature, with diverging strain fluctuations in the vicinity of the transition and a corresponding maximum of the apparent Poisson's ratio. Connecting with recent studies of mechanical criticality in athermal networks \cite{Broedersz2011, Sharma2016a, Feng2016, Shivers2018scaling, Rens2018, Merkel2019}, we demonstrate that this maximum occurs at a connectivity-controlled strain corresponding to macroscopic crossover between distinct mechanical regimes, with large-scale, collective network rearrangements as a branched, system-spanning network of tensile force chains develops at the transition. Our results highlight the influence of collective properties on the nonlinear mechanics of athermal networks and suggest that controlling connectivity could enable the design of tailored elastic anomalies in engineered fiber networks.

\begin{figure}[ht]
	\includegraphics[width=0.9\columnwidth]{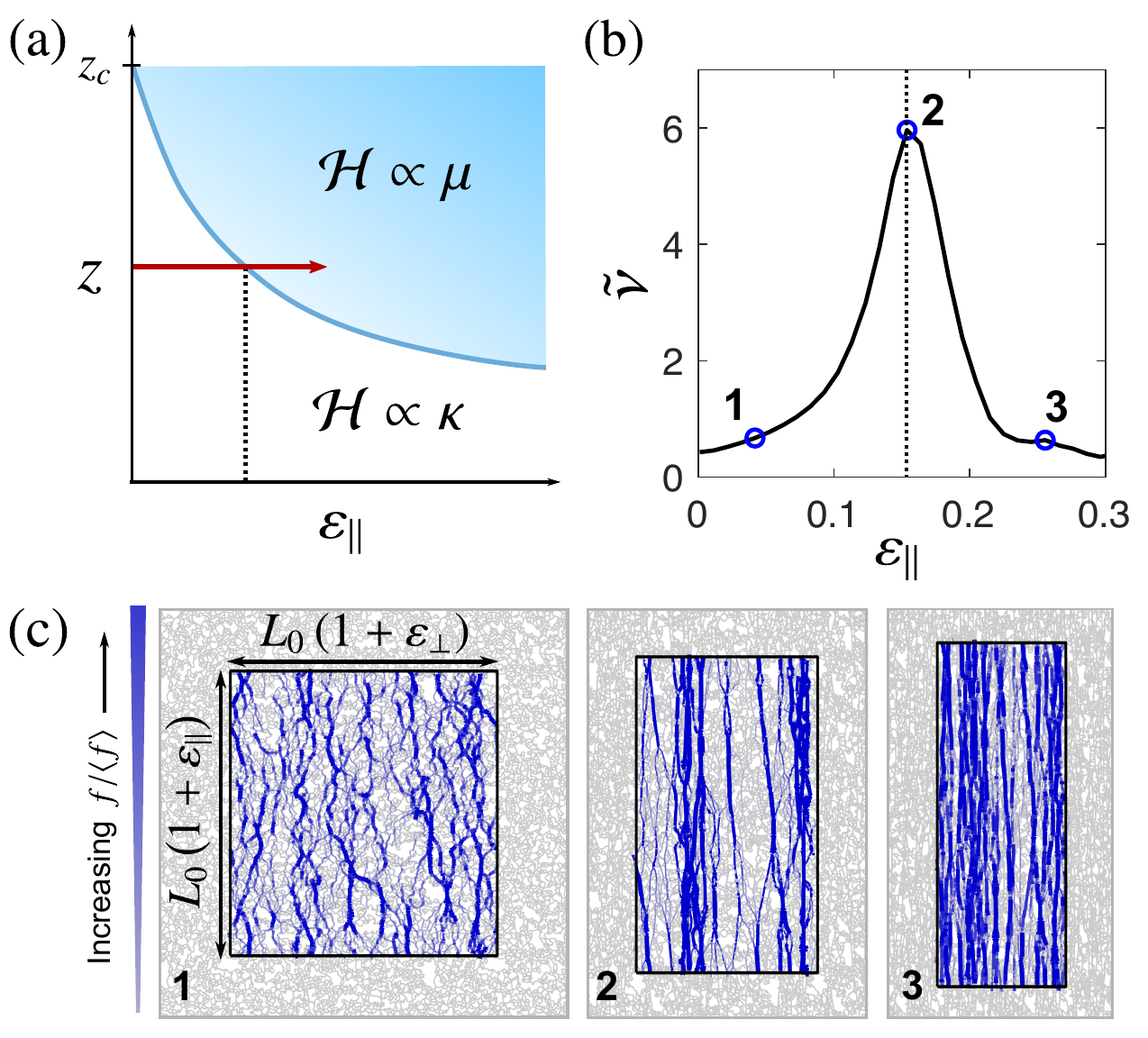}
	\caption{\label{fig1} (a) Under applied extensional strain $\varepsilon_\parallel$ (red arrow) with free transverse strains, subisostatic ($z < z_c$) athermal fiber networks transition from a soft, bending-dominated regime ($\mathcal{H}\propto\kappa$, floppy in the limit of $\kappa \to 0$) to a stiff, stretching-dominated regime ($\mathcal{H}\propto\mu$) at a critical applied strain $\varepsilon_{\parallel,c}$ (dotted line) that increases with decreasing $z$. As $z\to z_c$, $\varepsilon_{\parallel,c}\to0$.  (b) The incremental Poisson's ratio $\tilde{\nu}=-\partial\varepsilon_\perp/\partial\varepsilon_\parallel$ exhibits a peak at the critical strain, indicated by the dotted line. The black curve corresponds to a 2D packing-derived network with $\tilde{\kappa} = 10^{-5}$ and $z = 3.2$.  Network configurations corresponding to the numbered circles are shown in (c). Here, the black box represents the deformation of the initially square periodic boundaries. Bonds under greater tension $f$ than the average, $\langle f \rangle$, are colored blue with thickness proportional to $f/\langle f\rangle$.}
\end{figure}

Recent work has demonstrated that the strain-stiffening effect in crosslinked networks of stiff athermal semiflexible biopolymers, such as collagen, which can be modeled as elastic rods with bending modulus $\kappa$ and stretching modulus $\mu$, can be understood as a mechanical phase transition between a bending-dominated regime and a stretching-dominated regime at an applied shear or extensional strain governed by the average network coordination number $z$ \cite{Feng2016, Sharma2016a, Shivers2018scaling, Sheinman2012}. Despite being athermal, such networks exhibit classical signatures of criticality near this transition, including power-law scaling of the elastic moduli with strain and system-size-dependent nonaffine strain fluctuations indicative of a diverging correlation length \cite{Sharma2016a, Shivers2018scaling}.  In the limit of $\kappa\to0$, stiffening corresponds to the rearrangement of the network to form a marginally stable, highly heterogenous network of branched force chains \cite{Heussinger2007a, Shivers2019} similar to the force networks observed in marginal jammed packings under compressive or shear strain \cite{Radjai1996, OHern2001, Majmudar2005}. Prior work has considered this rigidity transition in networks under applied simple shear \cite{Sharma2016a, Shivers2018scaling, Rens2018, Merkel2019} or bulk strain \cite{Sheinman2012, Rens2018, Merkel2019}, with quantitative agreement between shear experiments on collagen and simulations \cite{Jansen2018}.

We find that an analogous collective mechanical phase transition controls the mechanics of networks under uniaxial strain with free orthogonal strains.  In athermal semiflexible polymer networks, strain-stiffening and the nonlinear Poisson effect occur at a critical extensional strain controlled by network connectivity, corresponding to a transition from a bending-dominated regime to one dominated by stretching. The expected phase diagram in connectivity-strain space is sketched in Fig.\ \ref{fig1}a. As applied strain drives a network to approach and cross the critical strain boundary, the network's mechanics become stretching-dominated and the resultant nonlinear strain-stiffening induces dramatic transverse contraction coinciding with a peak in the incremental Poisson's ratio $\tilde{\nu}$ (see Fig.\ \ref{fig1}b). Concurrent with this transition, the system exhibits nonaffine strain fluctuations which grow by orders of  magnitude as criticality is approached (either by decreasing $\kappa$ or approaching the critical strain). We demonstrate that this phenomenon occurs irrespective of the details of the underlying network structure, consistent with past observations of networks under simple shear \cite{Sharma2016a, Shivers2019}. Our results suggest that the dramatic nonlinear Poisson effect observed in collagen and fibrin gels is macroscopic evidence of this critical rigidity transition.

\textit{Models}---We consider 2D and 3D disordered networks comprising interconnected 1D Hookean springs with stretching modulus $\mu$, in addition to bending interactions with modulus $\kappa$ between selected adjacent bonds. To explore the influence of network structure on the strain-driven stiffening transition, we test a variety of network geometries, including Mikado networks \cite{Head2003}, 2D and 3D jammed packing-derived (PD) networks  \cite{Shivers2019}, 3D Voronoi networks \cite{Picu2018}, and 3D random geometric graph (RGG) networks \cite{Beroz2017}. Preparation of specific network structures is discussed in Supplementary Material. For an arbitrary network configuration, the total network Hamiltonian $\mathcal{H} = \mathcal{H}_s + \mathcal{H}_b$ consists of a stretching contribution,
\begin{equation} \label{eq_stretching}
\mathcal{H}_s = \frac{\mu}{2}\sum_{ij}{\frac{\left(\ell_{ij}-\ell_{ij,0}\right)^2}{\ell_{ij,0}}},
\end{equation} 
in which the sum is taken over connected node pairs $ij$, $\ell_{ij}$ is the length of the bond connecting nodes $i$ and $j$, and $\ell_{ij,0}$ is the corresponding rest length,
as well as a bending contribution,
\begin{equation} \label{eq_bending}
\mathcal{H}_b = \frac{\kappa}{2}\sum_{ijk}{\frac{\left(\theta_{ijk}-\theta_{ijk,0}\right)^2}{\ell_{ijk,0}}}.
\end{equation}
in which the sum is taken over connected node triplets $ijk$, $\theta_{ijk}$ is the angle between bonds $ij$ and $jk$, $\theta_{ijk,0}$ is the corresponding rest angle, and $\ell_{ijk,0} = (\ell_{ij,0}+\ell_{jk,0})/2$. 
For Mikado networks, which we designate to have freely hinging crosslinks, the sum in Eq. \ref{eq_bending} is taken only over consecutive node triplets along initially collinear bonds. Following prior work, we set $\mu = 1$ and vary the dimensionless bending rigidity $\tilde{\kappa} = \kappa/(\mu\ell_c^2)$ \cite{Licup2015,Shivers2019}, where $\ell_c$ is the average bond length. All network models utilize generalized Lees-Edwards periodic boundary conditions \cite{Lees1972, Shivers2019}, which specify that the displacement vectors between each network node and its periodic images transform according to the deformation gradient tensor $\bm{\Lambda}$, while the relative positions of nodes within the network are unconstrained. We consider purely extensional strain, with $\Lambda_{ii} = 1+\varepsilon_{i}$, where $\varepsilon_i$ is the strain along the $i$-axis relative to the initial configuration.
The normal stress components $\sigma_{ii}$ are computed as $\sigma_{ii}=(\partial\mathcal{H}/\partial\varepsilon_i)/V$, in which $V$ is the system's volume. Unless otherwise stated, all curves correspond to an average over 15 samples. 

To measure the nonlinear Poisson effect, we apply quasi-static longitudinal extensional strain $\varepsilon_{\parallel}\equiv\varepsilon_1$ in small increments $\delta\varepsilon_\parallel=\varepsilon_{\parallel,n}-\varepsilon_{\parallel,n-1}$ and, at a given strain, first allow the system to reach mechanical equilibrium by minimizing the network's Hamiltonian using the L-BFGS algorithm. After each extensional strain step, we simulate free transverse boundaries by incrementally varying the transverse strain(s) $\varepsilon_2$ (and $\varepsilon_3$ in 3D) in order to reduce the corresponding transverse normal stress component(s) to zero, i.e. $\left|\partial\mathcal{H}/\partial\varepsilon_i\right|\approx0$. In 2D the single transverse strain is $\varepsilon_{\perp}\equiv\varepsilon_2$, whereas in 3D the stresses along the two transverse axes are relaxed independently and we define the transverse strain, for the purposes of computing the incremental Poisson's ratio, as $\varepsilon_{\perp}\equiv(\varepsilon_2+\varepsilon_3)/2$. For orientationally isotropic network models, $\varepsilon_2$ and $\varepsilon_3$ are equivalent in the limit of large system size. The differential Young's modulus $\tilde{E}$ is computed as $\tilde{E} = \partial\sigma_\parallel/\partial\varepsilon_\parallel$.

\begin{figure}[ht]
	\includegraphics[width=0.9\columnwidth]{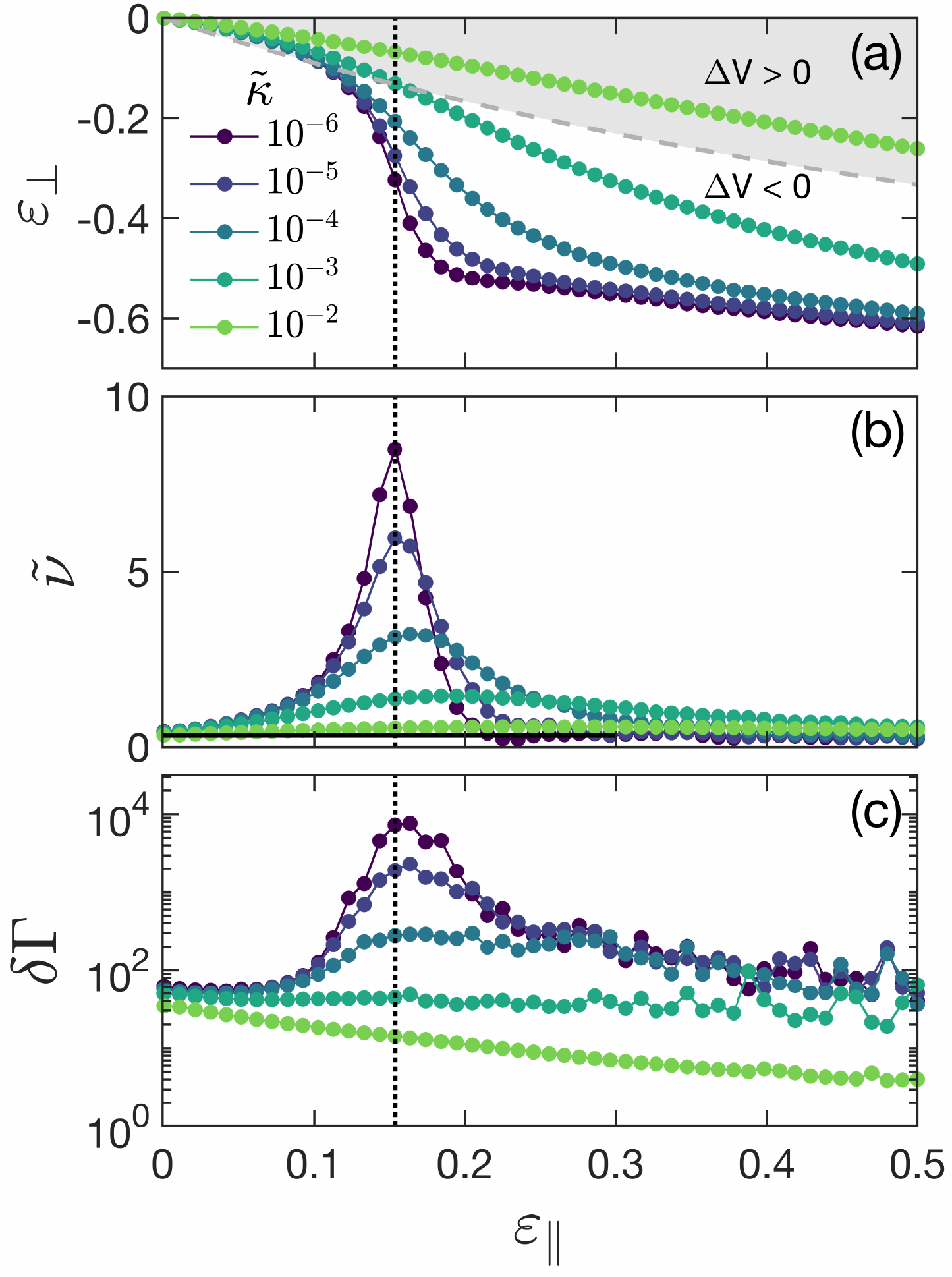}
	\caption{\label{fig2} (a) Relaxed transverse strain $\varepsilon_\perp$ as a function of applied extensional strain $\varepsilon_\parallel$ for 2D packing-derived networks with $z=3.2$ and varying $\tilde{\kappa}$. For large $\tilde{\kappa}$, networks deform linearly up to relatively large strains. The gray dashed line corresponds to constant volume, $\Delta V \equiv V - V_0 = 0$. In the limit of low $\tilde{\kappa}$, networks deform linearly at low strains, with a linear Poisson's ratio less than 1, but exhibit a significant increase in transverse contraction at a critical strain $\varepsilon_{\parallel,c}$, indicated by the dotted black line.  (b) The magnitude of the incremental Poisson's ratio $\tilde{\nu}=-\partial\varepsilon_\perp/\partial\varepsilon_\parallel$ peaks at the critical strain and increases with decreasing $\tilde{\kappa}$. (c) At the critical strain, we observe a corresponding peak in the nonaffine strain fluctuations $\delta\Gamma$ which increases in magnitude as $\tilde{\kappa}$ is decreased.}
\end{figure}

\textit{Results}---Subisostatic athermal networks undergo a transition from a bending-dominated regime to a stiff stretching-dominated regime at a critical applied shear or extensional strain \cite{Vahabi2016, Licup2016}. Recent work showed that athermal networks under extensional strain with free transverse strains, which we consider in this work, undergo a similar transition from a bending-dominated to stretching-dominated regime corresponding with strain-stiffening \cite{Picu2018}.  To examine the influence of bending rigidity on this transition, we first consider 2D packing-derived networks with fixed connectivity  $z = 3.2<z_c$ and varying reduced bending rigidity $\tilde{\kappa}$. In Fig.\ \ref{fig2}a, we plot the relaxed transverse strain $\varepsilon_{\perp}$ as a function of applied longitudinal extensional strain $\varepsilon_{\parallel}$, with the corresponding incremental Poisson's ratio $\tilde{\nu} = -\partial\varepsilon_\perp/\partial\varepsilon_\parallel$ shown in Fig.\ \ref{fig2}b. The fraction of the total network energy due to bending interactions $\mathcal{H}_b/\mathcal{H}$ as a function of strain is shown in Supplementary Material. Networks with high $\tilde{\kappa}$ deform approximately linearly up to relatively large applied strains, with minimal strain-dependence of $\tilde{\nu}$. In contrast, networks with low $\tilde{\kappa}$ exhibit similar linear deformation (with $\tilde{\nu} < 1$) in the limit of small applied strain, but under increasing applied strain these undergo a transition to a much stiffer stretching-dominated regime, resulting in significant transverse contraction and thus a very large apparent Poisson's ratio. At larger strains, within the stretching-dominated regime, the networks again deform approximately linearly with an incremental Poisson's ratio $\tilde{\nu}<1$.  The transition occurs at a critical applied extension $\varepsilon_c$, which we define as the strain corresponding to the inflection point in the $\varepsilon_{\perp}$ vs. $\varepsilon_{\parallel}$ curve as $\kappa\to0$.  By definition, this inflection point corresponds to a peak in the incremental Poisson's ratio $\tilde{\nu}$, which grows with decreasing $\tilde{\kappa}$.

This unusual nonlinear Poisson effect results from the asymmetric nonlinear mechanical behavior of these materials, i.e. the fact that they stiffen dramatically under extensional strain but not under compression \cite{Kabla2007b, Vader2009}. Compressing a semiflexible polymer network induces normal stresses proportional to the bending rigidity $\kappa$ of the constituent polymers, whereas sufficient extension induces normal stresses proportional to the polymer stretching modulus $\mu$ \cite{Vahabi2016}. An athermal network under uniaxial extension with \textit{fixed} transverse boundaries will exhibit an increase in the magnitude of its normal stresses from $\sigma_{ii}\propto\kappa$ to $\sigma_{ii}\propto \mu$ at the critical strain, both along the strain axis ($\sigma_{\parallel}$) and the transverse axes ($\sigma_{\perp}$). Relaxing the transverse boundaries to satisfy $\sigma_{\perp}=0$ requires sufficient contraction along the transverse axes to balance the stiff, inward-pointing normal stress contributions ($\propto \mu$) with softer, outward-pointing contributions ($\propto \kappa$). Transverse contraction also necessarily reduces the stretching-induced contributions by decreasing the length of the network along the transverse axes. The amount of transverse contraction required to achieve $\sigma_{\perp}=0$ thus increases with the amount of stiffening (e.g decreasing $\kappa$ relative to $\mu$  requires more transverse contraction at the critical strain).  Any compressible material that stiffens under extension, but not under compression, should exhibit a similar nonlinear Poisson effect.

Past work showed that athermal networks under applied shear strain exhibit diverging nonaffine strain fluctuations at the critical strain, in the limit of $\tilde{\kappa} \to 0$, indicative of a diverging correlation length \cite{Sharma2016a, Sharma2016, Shivers2018scaling}. Concurrent with the strain-driven transition in this work, we observe similarly large internal strain fluctuations. We use an analogous measure of the same strain fluctuations for the deformation gradient tensor $\bm{\Lambda}$ defined above.  For the $n$th strain step, the incremental applied extensional strain $\delta\varepsilon_\parallel=\varepsilon_{\parallel,n}-\varepsilon_{\parallel,n-1}$ and relaxation of the transverse strain(s) transforms the deformation gradient tensor (i.e. the extensional strains defining the periodic boundary conditions) from $\bm{\Lambda}_{n-1}$ to $\bm{\Lambda}_{n}$, with additional internal rearrangement of the network nodes to achieve force balance. We compute the resulting differential nonaffinity $\delta\Gamma$  as
\begin{equation}
\delta\Gamma = \frac{1}{\ell_c^2\left(\delta\varepsilon_\parallel\right)^2}\left\langle\left\Vert\delta\bm{\mathrm{u}}_i-\delta\bm{\mathrm{u}}_i^\mathrm{aff}\right\Vert^2\right\rangle
\end{equation}
in which the average is taken over all nodes $i$, $\ell_c$ is the initial average bond length, $\delta\bm{\mathrm{u}}_i=\bm{\mathrm{u}}_{i,n}-\bm{\mathrm{u}}_{i,n-1}$ is the actual displacement of node $i$ after the extensional strain step and transverse strain relaxation, and $\delta\bm{\mathrm{u}}_i^\mathrm{aff}$ is the displacement of node $i$ corresponding to an affine transformation from the previous configuration at strain state $\bm{\Lambda}_{n-1}$ to the new strain state $\bm{\Lambda}_{n}$. Consistent with prior work examining networks under shear strain \cite{Shivers2018scaling}, we find that increasing $\tilde{\kappa}$ results in increasingly affine deformation (decreasing $\delta\Gamma$), whereas in the low-$\tilde{\kappa}$ limit we observe a peak in $\delta\Gamma$ at the critical strain which grows with decreasing $\tilde{\kappa}$ (see Fig.\ \ref{fig2}c). 

\begin{figure}[ht]
	\includegraphics[width=1.0\columnwidth]{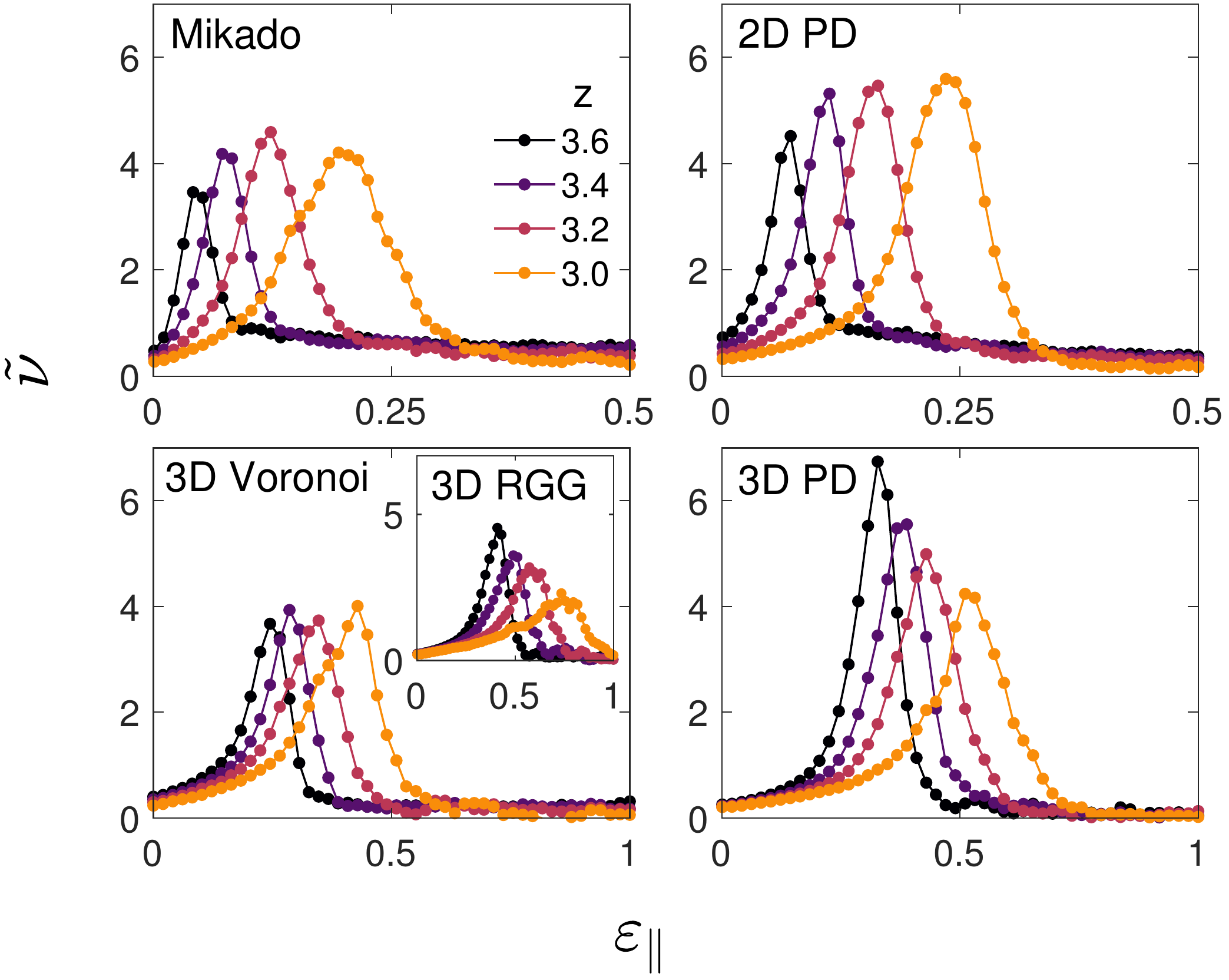}
	\caption{\label{fig3} Incremental Poisson's ratio $\tilde{\nu}=-\partial\varepsilon_\perp/\partial\varepsilon_\parallel$ as a function of applied extensional strain $\varepsilon_\parallel$ for various 2D and 3D network structures, as labeled in the top right of each panel, with varying connectivity $z$. The inflection point in each $\varepsilon_\perp$ vs. $\varepsilon_\parallel$ curve corresponds to a peak in $\tilde{\nu}$. The critical applied strain $\varepsilon_{\parallel,c}$ at which this inflection point occurs increases with decreasing $z$. }
\end{figure}

For athermal subisostatic networks under applied simple shear strain, the critical strain is governed by the average network connectivity $z$ \cite{Wyart2008, Sharma2016a, Sharma2016}, with the critical strain decreasing to zero as $z$ approaches the Maxwell isostatic value $z_c=2d$ , where $d$ is the dimensionality \cite{Maxwell1864}. As sketched in our hypothesized phase diagram (see Fig.\ \ref{fig1}a), we expect $z$ to similarly control the critical strain for networks under extensional strain with free orthogonal strains. In Fig.\ \ref{fig3}, we plot the incremental Poisson ratio $\tilde{\nu}$ as a function of $\varepsilon_\parallel$ for several network geometries in 2D and 3D with varying $z$. While the precise location of the critical strain for a given connectivity is sensitive to the choice of network structure, we find that all networks tested exhibit behavior that is qualitatively consistent with the proposed phase diagram, with a critical strain $\varepsilon_{\parallel,c}$ that decreases as $z\to z_c$.

\begin{figure}[ht]
	\includegraphics[width=0.9\columnwidth]{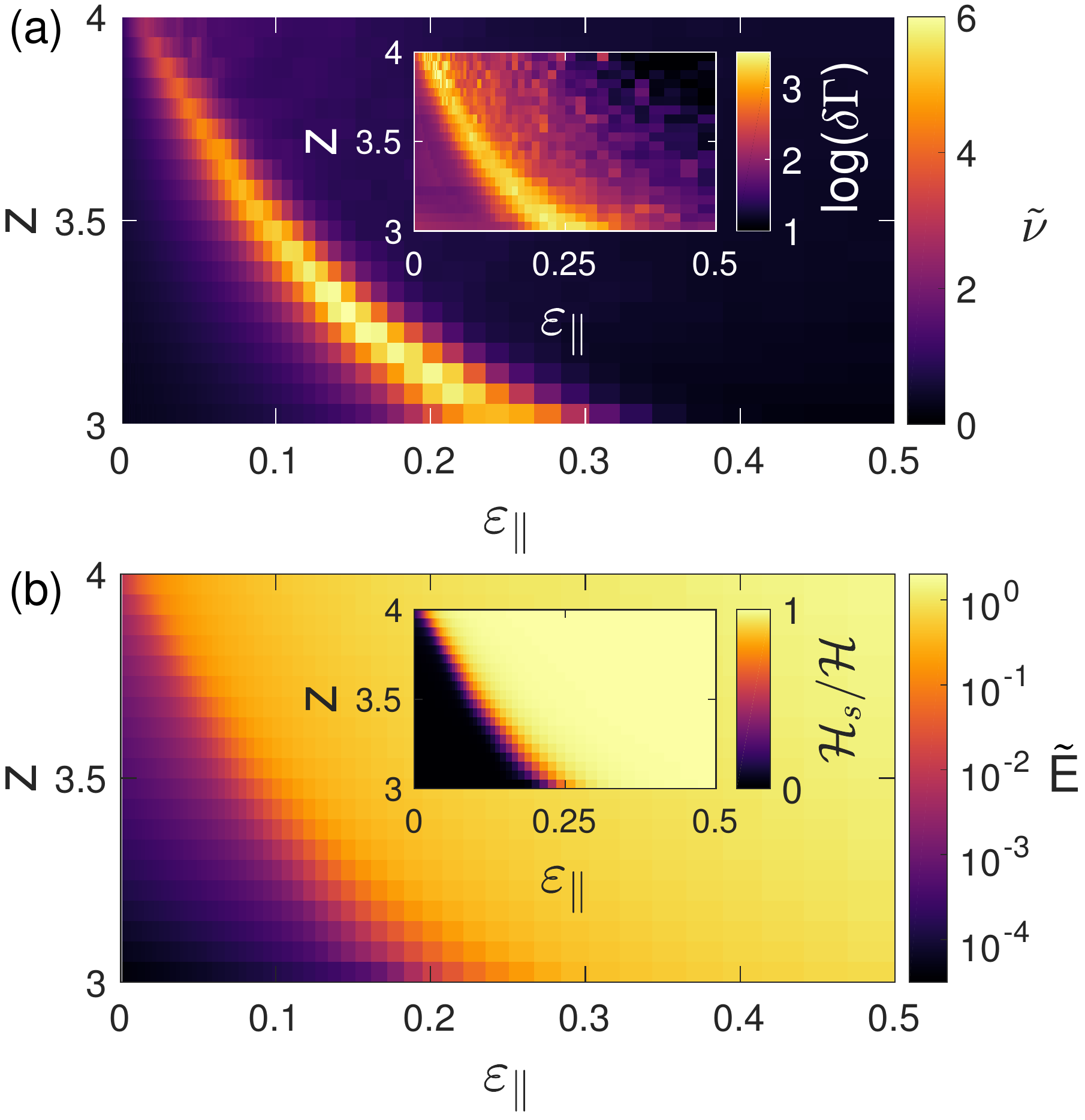}
	\caption{\label{fig4} (a) Incremental Poisson's ratio $\tilde{\nu} = -\partial\varepsilon_\perp/\partial\varepsilon_\parallel$ as a function of applied extensional strain $\varepsilon_\parallel$ and average network coordination $z$ for 2D packing-derived networks with $W = 100$ and $\tilde{\kappa} = 10^{-5}$. Inset: For a given network coordination $z$, the differential nonaffinity $\delta\Gamma$ shows a peak corresponding to the peak in the incremental Poisson ratio $\tilde{\nu}$ at the same critical applied extension. (b) Differential Young's modulus $\tilde{E} = \partial\sigma_{\parallel}/\partial\varepsilon_\parallel$ for the same networks as in (a). Inset: Corresponding stretching energy fraction $\mathcal{H}_s/\mathcal{H}$.}
\end{figure}

We also explicitly map out a phase diagram for packing-derived networks in 2D. In Fig.\ \ref{fig4}a, we plot both the incremental Poisson's ratio $\tilde{\nu}$ and differential nonaffinity $\delta\Gamma$ for 2D PD networks as a function of applied strain over a range of $z$ values up to the 2D isostatic point, $z_c = 4$. Both quantities become maximal at a critical strain that approaches 0 as $z\to z_c$. We plot the corresponding differential Young's modulus $\tilde{E} = \partial\sigma_{\parallel}/\partial\varepsilon_\parallel$ as a function of $z$ and $\varepsilon_\parallel$ in Fig.\ \ref{fig4}b, demonstrating that the transition of the network from the soft, bending-dominated regime ($\tilde{E}\propto\kappa$) to the stiffer, stretching-dominated regime ($\tilde{E}\propto\mu$) coincides with peaks in both the incremental Poisson's ratio and the differential nonaffinity (Fig.\ \ref{fig4}a). Further, we find that the differential Young's modulus scales as a power law $\tilde{E}\propto\left|\varepsilon_\parallel - \varepsilon_{\parallel,c}\right|^f$ above the critical strain (see Supplementary Material).

\textit{Discussion}---We have demonstrated that the nonlinear Poisson effect observed in subisostatic networks is a direct consequence of a strain-driven collective mechanical phase transition. Whereas the large apparent Poisson's ratios observed in such networks at finite strains can be qualitatively understood as resulting from their highly asymmetric mechanical properties, i.e. that they stiffen dramatically under finite extension but remain comparatively soft under compression, as discussed conceptually in Refs.\ \cite{Kabla2007b,Vader2009}, we have demonstrated that this asymmetry becomes maximized at a critical phase boundary controlled by strain and connectivity. At this boundary, a network exhibits diverging strain fluctuations as it collectively rearranges to transition from a soft, bending-dominated regime to a stiff, stretching-dominated regime. In the latter, marginally stable state, the mechanics become dominated by an underlying branched network of bonds under tension, which generates tensile transverse normal stresses that drive the lateral contraction of the sample against the weaker compressive stresses. 
This results in an apparent Poisson's ratio that exceeds $1/2$ at the phase transition and grows as a function of the relative magnitude the stiff and soft contributions, $\mu/\kappa$.  

Using simulations of a variety of network architectures in 2D and 3D, we have shown that this effect is robustly controlled by connectivity and occurs independently of the precise underlying network structure. Further, we have demonstrated critical scaling of the differential Young's modulus (see Supplementary Material) similar to what has been shown for the shear modulus of collagen networks \cite{Sharma2016a}. This suggests that experimental measurements of the differential Young's modulus of collagen gels under uniaxial strain should quantitatively fit the predicted scaling form, with a given sample exhibiting a peak in the incremental Poisson's ratio at the transition point. Further work could enable prediction of the local stiffness in the extracellular matrix based on the observed local strain asymmetry.

\begin{acknowledgments}
This work was supported in part by the National Science Foundation Division of Materials Research (Grant DMR-1826623) and the National Science Foundation Center for Theoretical Biological Physics (Grant PHY-1427654). J.L.S. acknowledges additional support from the Ken Kennedy Institute Graduate Fellowship and the Riki Kobayashi Fellowship in Chemical Engineering. 
\end{acknowledgments}

%

\section{Supplemental Material}

\subsection{I. Network generation}

Mikado networks \cite{Wilhelm2003,Head2003} are prepared by depositing fibers of length $L$ with random locations and orientations into a 2D periodic square unit cell of side length $W$ and adding freely hinging crosslinks at all fiber intersections. We use $L = 4$ and $W = 30$ and continue depositing fibers until the average coordination number, after the removal of dangling ends, is $z \approx 3.6$. This yields an average crosslink density of $L/\ell_c\approx 11$, where $\ell_c$ is the average distance between crosslinks. We then randomly remove bonds and dangling ends until the desired $z$ is reached. We impose a minimum segment length $\ell_{min} = W/1000$. 

2D and 3D packing-derived networks are prepared as in prior work \cite{Shivers2019}. For 2D PD networks, we randomly place $N = W^2$ radially bidisperse disks with $r\in\{r_0,\phi r_0\}$ in a periodic square unit cell of side length $W$ and incrementally increase $r_0$ from $0$, allowing the system to relax at each step, until the packing becomes isostatic and develops a finite bulk modulus. We use $\phi = 1.4$ to avoid long-range crystalline order \cite{Koeze2016}. At this point, we generate a contact network between overlapping disks.  For 3D PD networks, we follow the same procedure beginning with $N=W^3$ radially bidisperse spheres, also with $r\in\{r_0,\phi r_0\}$ and $\phi = 1.4$, in a periodic cubic unit cell of side length $W$. We use $W=20$ in 3D and $W=100$ in 2D. For sufficiently large systems, this yields a contact network with $z\approx2d$ \cite{VanHecke2010,Ohern2003,Dagois-Bohy2012}. After generating the initial network, we randomly remove bonds and dangling ends until the desired $z$ is reached.

We generate 3D Voronoi networks by randomly distributing $N$ seed points in a periodic cubic unit cell of side length $W$, from which we generate a Voronoi diagram using the CGAL library \cite{cgal}. We choose $N$ so that the final network will have roughly $W^3$ nodes for consistency with our other 3D models, and use $W = 15$. These have initial average coordination $z = 4$.

Three-dimensional random geometric graph (RGG) models have been shown to capture the micromechanics of collagen and fiber networks \cite{Beroz2017}. Following Ref.  \cite{Beroz2017}, we generate RGGs of $N=W^3$ vertices in a periodic box of side length $W$, where each pair of nodes is connected with probability $P_c\propto e^{-\ell/L}/\ell^2$, where $\ell$ is the distance between two vertices and $L=1$ is the length scale of a typical bond. We impose a minimum bond length of $\ell_{min}=0.5$. We generate RGG networks with $W = 20$ and initial average coordination number $z = 4$, which we further dilute to the desired $z$ by randomly removing bonds and dangling ends.

\subsection{II. Midpoints}

\begin{figure}[!htb]
	\includegraphics[width=1.0\columnwidth]{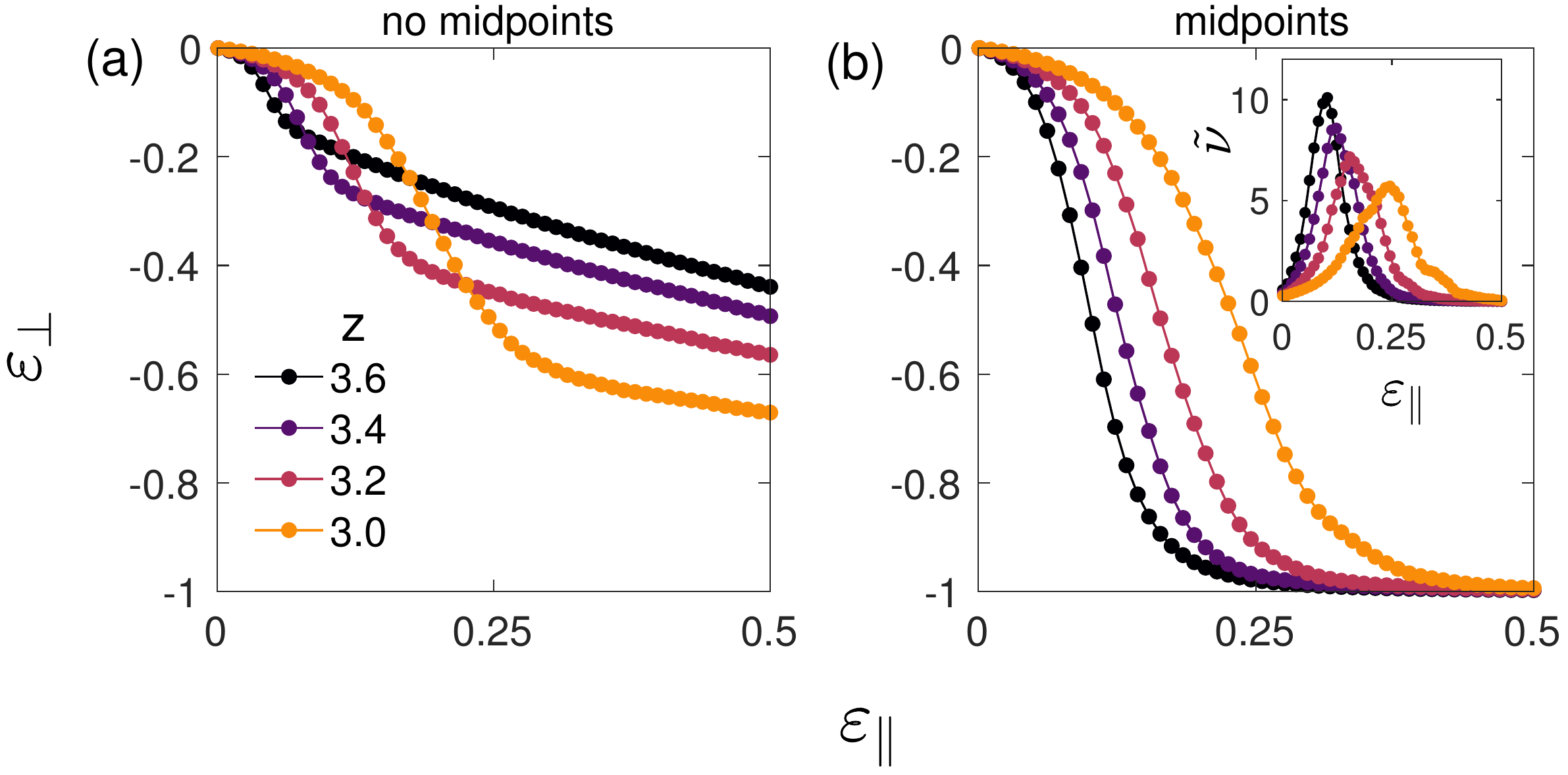}
	\caption{\label{figS1} Transverse strain $\varepsilon_\perp$ as a function of applied extensional strain $\varepsilon_\parallel$ for Mikado networks with $\tilde{\kappa} = 10^{-5}$ and varying connectivity $z$ (a) without buckling of individual bonds and (b) with buckling of individual bonds, in which an extra node is added at the midpoint of each bond. Inset: Incremental Poisson's ratio $\tilde{\nu}$ as a function of $\varepsilon_\parallel$ for networks with midpoints. The corresponding plot for Mikado networks without midpoints is shown in Fig. 3a.}
\end{figure}

The amount of contraction induced by the onset of stiffening depends on the compressibility of the network. Mikado and 2D PD networks with relatively high values of $z$ exhibit less dramatic contraction upon transitioning to the stretching-dominated regime than the same network types with lower $z$. This is a result of the fact that more dilute (lower $z$) networks have fewer highly connected regions within them and are thus less resistant to compression than more highly connected networks. To verify that this is the case, in Fig.\ \ref{figS1}, we consider the same Mikado structures as in Fig.\ 3, in which we have now added a midpoint hinge to each bond in order to allow the buckling of individual bonds with the same bending energy penalty as is used between adjacent bonds. While the location of the transition in strain is still controlled by connectivity, networks with midpoints contract more dramatically at the critical strain than those without midpoints (for fixed $\tilde{\kappa}$), with larger corresponding peaks in $\tilde{\nu}$. This is because the normal stress induced by transverse compression in networks with midpoints is strictly proportional to $\kappa$, whereas in networks without midpoints there can be additional, stronger contributions ($\propto\mu$) from locally stiff regions at large compression levels.

 \FloatBarrier
 
\subsection{III. Energy contributions for networks with fixed $z$ and varying $\tilde{\kappa}$}

\begin{figure}[!htb]
	\includegraphics[width=0.8\columnwidth]{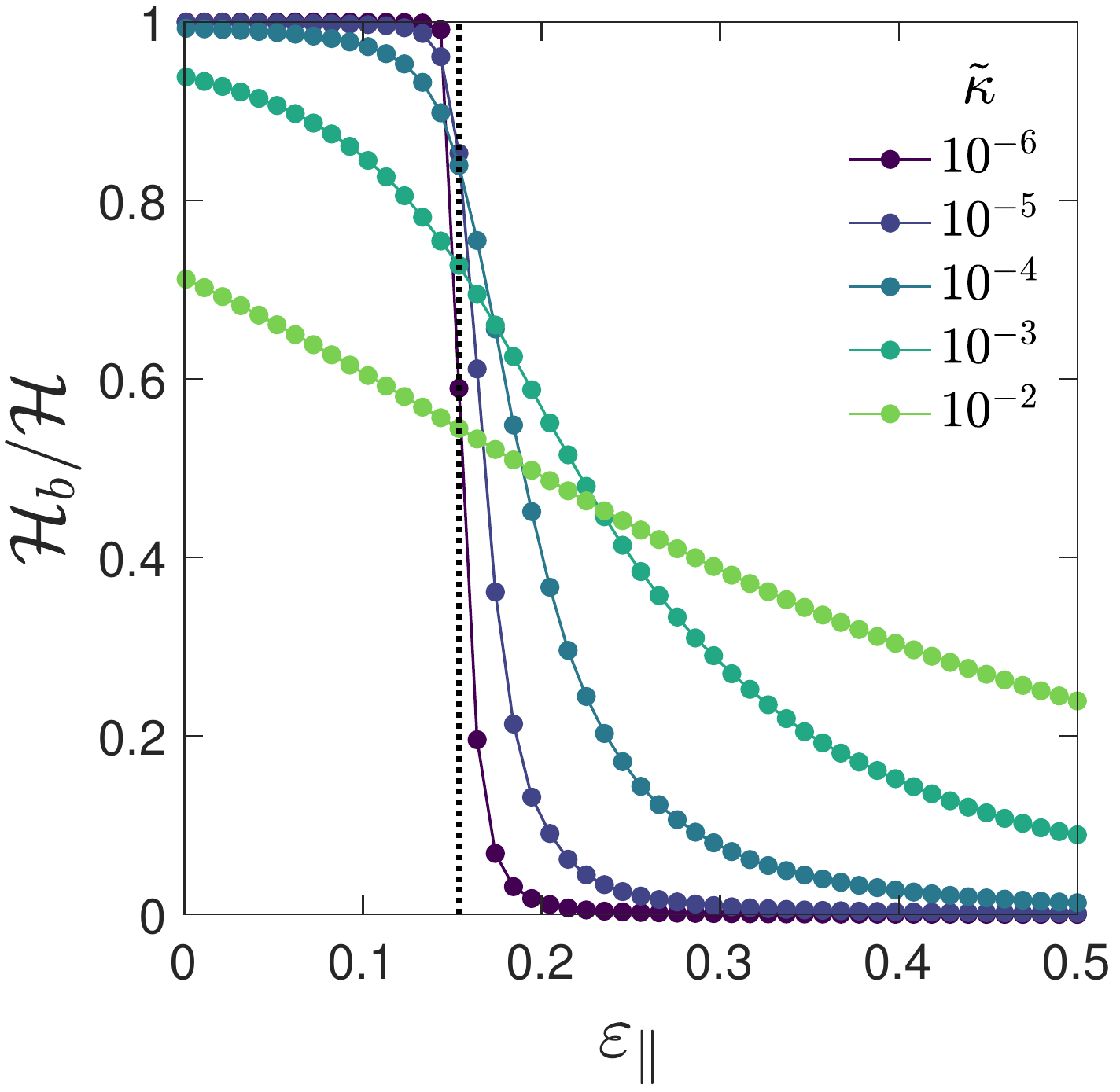}
	\caption{\label{figS2}  Bending energy fraction $\mathcal{H}_b/\mathcal{H}$ for a 2D packing-derived network with $z=3.2$, $W = 100$ and varying $\tilde{\kappa}$. The dotted line represents the critical strain $\varepsilon_{\parallel,c}$ corresponding to the peak in the incremental Poisson's ratio. }
\end{figure}

In Fig. \ref{figS2}, we plot the bending energy fraction for a 2D packing-derived network with varying $\tilde{\kappa}$. In the low-$\tilde{\kappa}$ limit, we observe a clear transition from a bending-dominated to stretching-dominated regime at the critical strain, corresponding to the maximum value of the incremental Poisson's ratio $\tilde{\mu}$.

 \FloatBarrier
 
\subsection{IV. Critical scaling}

\begin{figure}[!htb]
	\includegraphics[width=0.8\columnwidth]{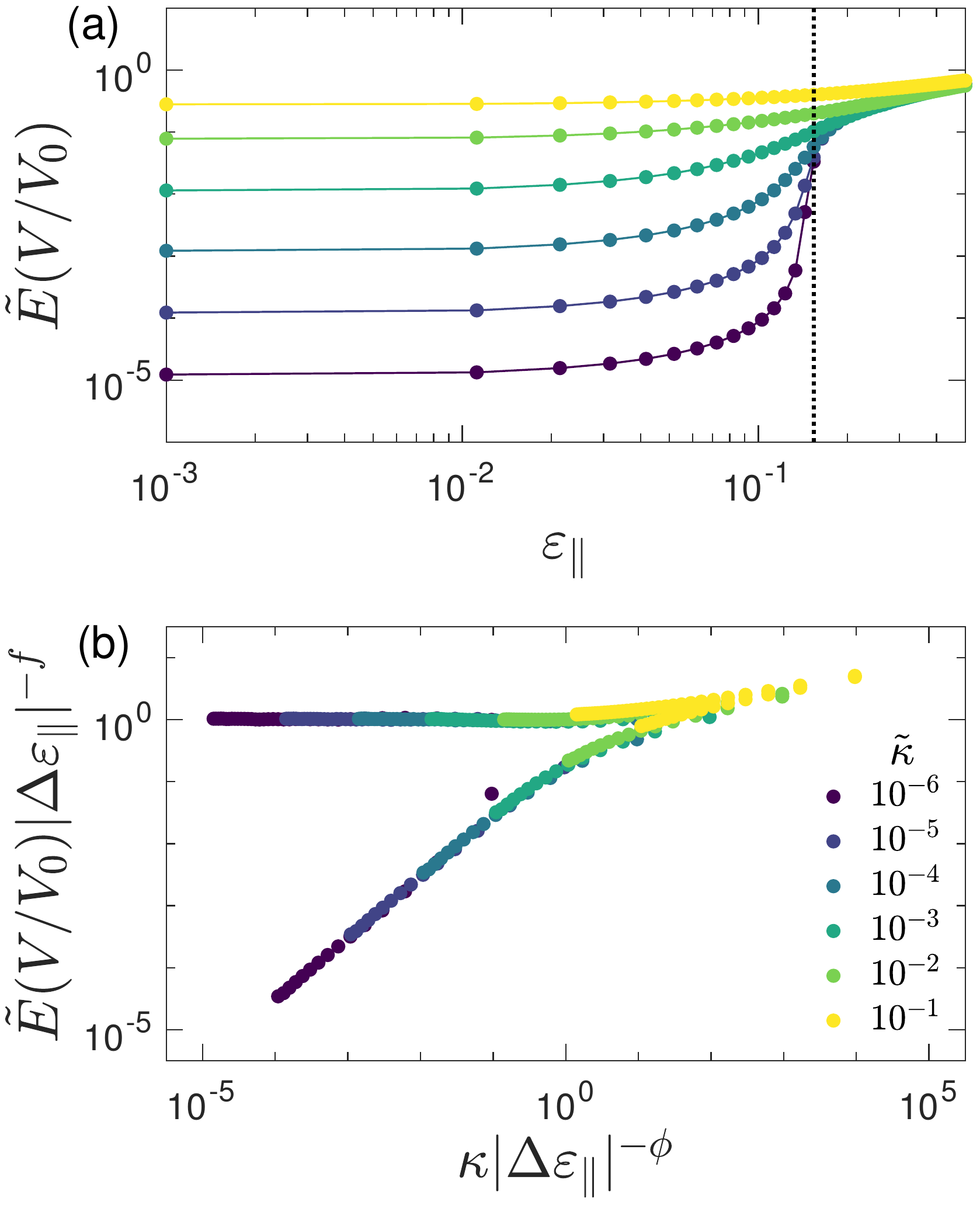}
	\caption{\label{figS3} (a) Stiffening curves for a 2D packing-derived network with varying $\kappa$, $z = 3.2$, $W = 140$. (b) Collapse of the curves in (a) according to Eq. \ref{scaling} with exponents $f = 0.55$ and $\phi = 2.5$. }
\end{figure}

For subisostatic networks in the limit of small applied strain $\varepsilon_\parallel$, the differential Young's modulus $\tilde{E}=\partial\sigma_\parallel/\partial\varepsilon_\parallel$ is proportional to the bending rigidity $\tilde{\kappa}$. Above the critical applied strain $\varepsilon_{\parallel,c}$, $\tilde{E}$ is independent of $\tilde{\kappa}$ and scales as a power law with respect to the distance (along the strain axis) to the critical strain, i.e. $\tilde{E}\propto|\Delta\varepsilon_\parallel|^f$, where $\Delta\varepsilon_\parallel = \varepsilon_\parallel - \varepsilon_{\parallel,c}$. Following prior work \cite{Sharma2016a}, we can capture both regimes with the scaling form
\begin{equation} \label{scaling}
\tilde{E}\frac{V}{V_0}\propto|\Delta\varepsilon_\parallel|^f\mathcal{G}_\pm\left(\frac{\tilde{\kappa}}{|\Delta\varepsilon_\parallel|^\phi}\right)
\end{equation}
in which the scaling function $\mathcal{G}_\pm$ has branches corresponding to positive and negative values of $\Delta\varepsilon_\parallel$, and the factor $V/V_0$ corrects for the apparent change in $\tilde{E}$ due to the change in the system's volume $V$ from the initial volume $V_0$.  In Fig.\ \ref{figS3}a, we show stiffening curves for a large packing-derived network with varying $\tilde{\kappa}$ and demonstrate scaling collapse according to the above scaling form with $f = 0.55$ and $\phi = 2.5$ (see Fig. \ref{figS3}b).

\begin{figure}[!htb]
	\includegraphics[width=0.8\columnwidth]{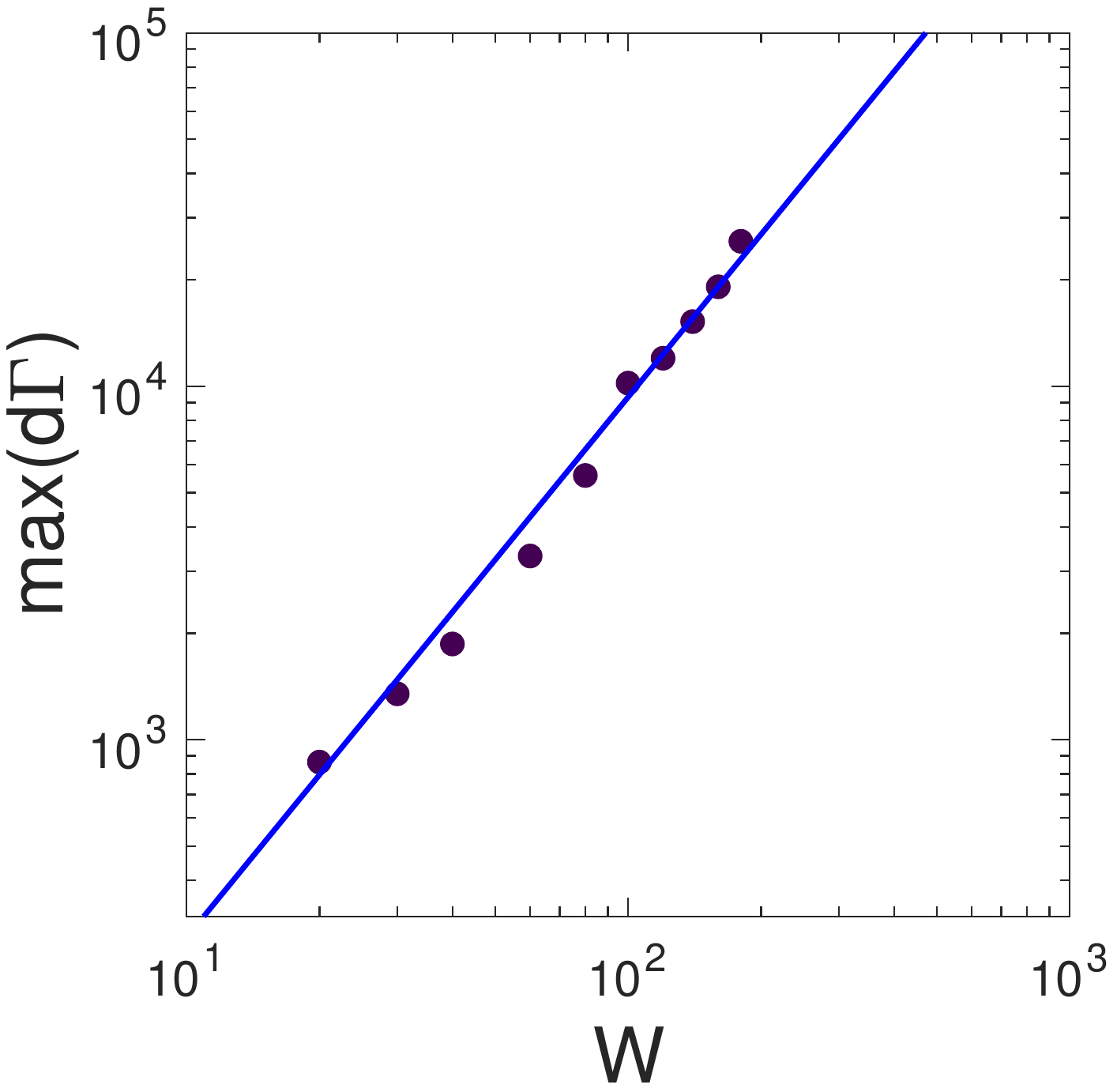}
	\caption{\label{figS4}  In the low-$\tilde{\kappa}$ limit, the magnitude of the differential nonaffinity at the critical strain, $\rm{max}(\delta\Gamma)$, grows with increasing system size $W$ as $\mathrm{max}(\delta\Gamma)\propto W^{(\phi-f)/\nu}$ (solid line), with $f = 0.55$ and $\phi = 2.5$ obtained from the scaling collapse of the differential Young's modulus and $\nu = (f + 2)/d$ with dimensionality $d=2$, as derived in Ref. \cite{Shivers2018scaling}. These data correspond to averages over 2D packing-derived networks with $z = 3.2$ and $\tilde{\kappa} = 10^{-6}$.}
\end{figure}

Recent work described a hyperscaling relation between $f$, the dimensionality $d$, and the correlation length exponent $\nu$,  \cite{Shivers2018scaling}
\begin{equation} \label{hyperscaling}
f = d\nu - 2,
\end{equation}
and predicted that, at the critical strain, the differential nonaffinity should scale with system size $W$ as 
\begin{equation}
\mathrm{max}(\delta\Gamma)\propto W^{(\phi-f)/\nu}.
\end{equation}
 Determining $\nu$ from Eq. \ref{hyperscaling} using $d = 2$ and $f = 0.55$ (from the prior scaling collapse of $\tilde{E}$), we observe good agreement between measured values of $\rm{max}(\delta\Gamma)$ and the predicted scaling of the differential nonaffinity with system size (see Fig. \ref{figS4}).
 
 \FloatBarrier

\end{document}